\documentclass[12pt]{article}
\usepackage{amssymb, amsmath}
\usepackage{amsfonts}
\usepackage{color}
\usepackage{cite}

\newcommand{\Section}[1]%
{\section{#1}\setcounter{equation}{0}%
\setcounter{theorem}{0}}

\def\re{\mathbb{R}}
\def\co{\mathbb{C}}

\def\ze{\mathbb{Z}}

{\par\noindent{\em #1:\ }}%
{~\rule{2mm}{2mm}\par\bigskip}

\begin{document}
\newpage\thispagestyle{empty}
{\topskip 2cm
\begin{center}
{\Large\bf A Concrete Example of Fractional Chern Insulator\\} 
\bigskip\bigskip
{\Large Tohru Koma\footnote{\small \it Department of Physics, Gakushuin University (retired), 
Mejiro, Toshima-ku, Tokyo 171-8588, JAPAN}
\\}
\end{center}
\vfil
\noindent
{\bf Abstract:} We present a concrete example of fractional Chern insulator whose fermion Hamiltonian 
consists of hopping and Coulomb repulsive interaction terms. Both of them are of finite range 
on the square lattice. In a strong coupling limit for the interaction Hamiltonian, 
we show that the Hall conductance is fractionally quantized to $1/2$ in the sense of the expectation 
value with respect to the four-fold degenerate ground state at the filling $3/8$. 
We also present a slightly different example in which there appears a long-range order of charge density wave with 
the Chern number $1$. 
\par
\noindent
\bigskip
\hrule
\bigskip\bigskip\bigskip
\vfil}

\Section{Introduction}

As is well known, the fractional quantum Hall effect occurs due to strong interactions between electrons 
at a fractional filling of a Landau level in a strong magnetic field in two dimensions. 
The strong interactions open a gap in the degenerate Landau level when the filling factor is fractional \cite{Komagap}, 
and the Hall conductance is quantized to the fractional value \cite{Komacondc}. 

This quantization is robust against generic perturbations, 
and can be explained from the topological point of view \cite{NTW,Koma1}. 
In this argument, the many-body ground state of the Hamiltonian must have a non-trivial degeneracy 
for the fractional quantization of the Hall conductance. In other words, a unique many-body ground state 
always shows an integer-valued Hall conductance. 

The fractional Chern insulators(FCI) as an interacting lattice fermion are an analogue of the fractional quantum Hall system.   
Naively, there seems to be only the difference between lattice and continuum fermions. 
However, there have been done many attempts to realize a fractional Chern insulator 
as a lattice fermion system \cite{RB,BL,PRS,LB,OM,LVP}. 
Of course, the treatment of strongly interacting fermions is not so easy. 

In the preset paper, we present a concrete example of the fractional Chern insulator whose Hamiltonian 
consists of hopping and Coulomb repulsive interaction terms. Both of them are of finite range 
on the square lattice. In a strong coupling limit for the Coulomb repulsion, 
we show that the ground state is four-fold degenerate, and that 
the Hall conductance is fractionally quantized to $1/2$ in the sense of the expectation 
value in the sector of the ground state at the filling $3/8$. 
We also present a slightly different example in which there appears a long-range order of charge density wave(CDW) with 
the Chern number $1$. Therefore, the two FCI and CDW phases can be clearly distinguished by the Chern numbers. 
 
The present paper is organized as follows: In Sec.~\ref{Sec:Model}, as a preliminary step, 
we describe models for fractional Chern insulators in a slightly generic setting. 
Our interaction Hamiltonian is given in Sec.~\ref{Sec:Hint}. In Sec.~\ref{Sec:hopHam}, we present our hopping Hamiltonian, 
and show our main results.  
In Sec.~\ref{Sec:CDWFCI}, we discuss the difference between the FCI and CDW phases. 
Appendix~\ref{HKmodel} is devoted to a short review of Hatsugai-Kohmoto model \cite{HK} whose Hamiltonian is used 
as a hopping Hamiltonian in the present paper.

\Section{A generic setting}
\label{Sec:Model} 

In order to clarify our aim of the present paper, we first consider a slightly generic setting 
about a class of lattice models which are believed to describe fractional Chern insulators \cite{RB,BL,PRS,LB}.  

Let $\Lambda$ be a subset of the square lattice $\ze^2$, and write $x=(x^{(1)},x^{(2)}) \in\ze^2$ 
for the site $x$ in the lattice $\Lambda$. In general, the Hamiltonian on the lattice $\Lambda$ is given by 
\begin{equation}
\label{geneHam}
H^{(\Lambda)}=\sum_{x,y\in\Lambda}t_{x,y}a_x^\dagger a_y +\sum_{x,y\in\Lambda:x\ne y}V_{x,y}n_xn_y,
\end{equation}
where $a_x^\dagger$ and $a_x$ are, respectively, the creation and annihilation fermion operators, which obey 
the anti-commutation relations, 
$$
\{a_x,a_y^\dagger\}=\delta_{x,y},\quad \mbox{and}\quad \{a_x,a_y\}=0,
$$
and $n_x:=a_x^\dagger a_x$; the hopping amplitudes $t_{x,y}\in\co$ satisfy the hermitian conditions $t_{y,x}=(t_{x,y})^\ast$, 
and $V_{x,y}\in\re$. We assume that the hopping and interaction Hamiltonians are of finite range. 
Namely, we will not consider a mean-field type model.  

In order to realize a fractional Chern insulator, we need to control suitably 
the model parameters in the Hamiltonian of (\ref{geneHam}). Mimicking the structure of the Landau band in 
the fractional quantum Hall system, the flat band structures \cite{SGKS} have been often used \cite{RB,BL,PRS} 
for the hopping Hamiltonian.   

However, intuitively, a strong Coulomb repulsion can be expected to induce a different configuration of the fermions from 
that for the band determined by the hopping Hamiltonian at a fractional filling.  
Therefore, it does not look like that a flat band structure is crucial for the fractional Chern insulators. 
Rather, we should control the Coulomb repulsion to realize a fractional Chern insulator. 
This is our motivation in the present paper. More precisely, 
in order to bring out the effect of the Coulomb interaction, we take the strong coupling limit. 

We should also remark that there are different recent approaches \cite{OM,LVP} to FCI from the above.

\Section{The present interaction Hamiltonian} 
\label{Sec:Hint}

In this section, we specify our interacting Hamiltonian, and discuss the structure of 
the degenerate ground state of the interaction Hamiltonian with no hopping Hamiltonian. 

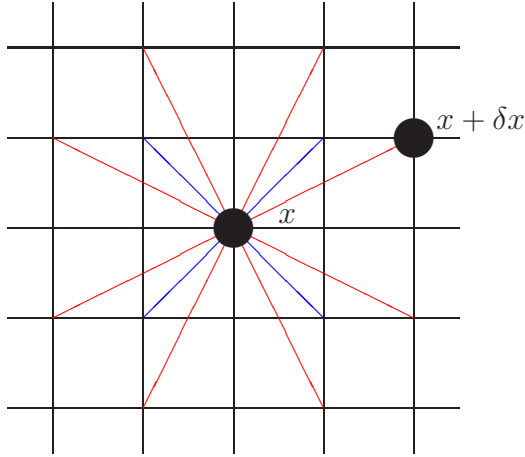
\begin{figure}[ht]
\setlength{\unitlength}{1.2cm}
\begin{center}
\begin{picture}(5,5)(0,0)
\put(0,0.5){\line(1,0){5}}
\put(0,1.5){\line(1,0){5}}
\put(0,2.5){\line(1,0){5}}
\put(0,3.5){\line(1,0){5}}
\put(0,4.5){\line(1,0){5}}
\put(0.5,0){\line(0,1){5}}
\put(1.5,0){\line(0,1){5}}
\put(2.5,0){\line(0,1){5}}
\put(3.5,0){\line(0,1){5}}
\put(4.5,0){\line(0,1){5}}
\put(2.5,2.5){\color{blue}\line(1,1){1}}
\put(2.5,2.5){\color{blue}\line(1,-1){1}}
\put(2.5,2.5){\color{blue}\line(-1,1){1}}
\put(2.5,2.5){\color{blue}\line(-1,-1){1}}
\put(2.5,2.5){\color{red}\line(1,2){1}}
\put(2.5,2.5){\color{red}\line(-1,2){1}}
\put(2.5,2.5){\color{red}\line(2,1){2}}
\put(2.5,2.5){\color{red}\line(2,-1){2}}
\put(2.5,2.5){\color{red}\line(1,-2){1}}
\put(2.5,2.5){\color{red}\line(-1,-2){1}}
\put(2.5,2.5){\color{red}\line(-2,1){2}}
\put(2.5,2.5){\color{red}\line(-2,-1){2}}
\put(3,2.55){$x$}
\put(4.75,3.6){$x+\delta x$}
\put(2.5,2.5){\circle*{0.8}}
\put(4.5,3.5){\circle*{0.8}}
\end{picture}
\end{center}
\caption{The red lines indicate the interactions of the second sum in the right-hand side of (\ref{Hint1}), 
and the blue lines indicate the interactions given by the Hamiltonian $H_{{\rm int},2}$ of (\ref{Hint2}).}
\label{interaction1}
\end{figure}

For this purpose, we introduce three sets of vectors as follows: 
$$
\mathcal{U}_1:=\{\pm e_1, \pm e_2 \}, \quad \mathcal{U}_2:=\{\pm e_1\pm e_2\}\quad \mbox{and} \quad 
\mathcal{U}_3:=\{\pm e_1\pm 2e_2, \pm e_2\pm 2e_1\},
$$
where $e_i\in\ze^2$ is the unit vector whose the $i$-th component is $1$ for $i=1,2$, i.e., $e_1=(1,0)$ and $e_2=(0,1)$. 
We choose the interaction Hamiltonian as follows: 
\begin{equation}
\label{Hint}
H_{\rm int}^{(\Lambda)}:=g_1H_{{\rm int}, 1}^{(\Lambda)}+g_2H_{{\rm int}, 2}^{(\Lambda)},
\end{equation}
where $g_1>0$ and $g_2>0$ are the coupling constants, and the two Coulomb interactions are given by 
\begin{equation}
\label{Hint1}
H_{{\rm int}, 1}^{(\Lambda)}:=\frac{1}{2}\sum_{x\in\Lambda}\sum_{\delta x\in\mathcal{U}_1}n_xn_{x+\delta x}
+\frac{1}{2}\sum_{x\in\Lambda}\sum_{\delta x\in\mathcal{U}_3}n_x n_{x+\delta x}
\end{equation}
and 
\begin{equation}
\label{Hint2}
H_{{\rm int}, 2}^{(\Lambda)}:=\frac{1}{2}\sum_{x\in\Lambda}\sum_{\delta x\in\mathcal{U}_2}n_xn_{x+\delta x}. 
\end{equation}
The first sum in the right-hand side of (\ref{Hint1}) is the usual nearest-neighbor interactions. 
The interactions given by the second sum in the right-hand side of (\ref{Hint1}) 
are depicted by the red lines in Fig.~\ref{interaction1}. The interactions of $H_{{\rm int},2}^{(\Lambda)}$ of (\ref{Hint2}) 
are also depicted by the blue lines in Fig.~\ref{interaction1}.

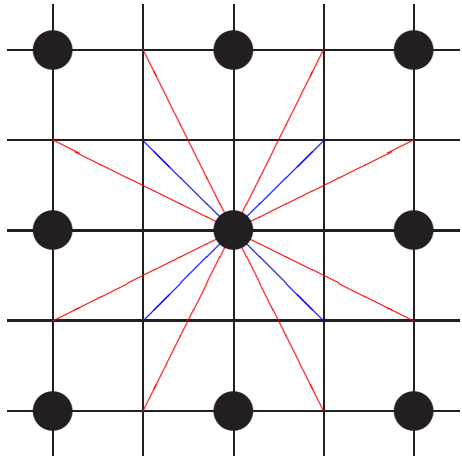
\begin{figure}[ht]
\setlength{\unitlength}{1.2cm}
\begin{center}
\begin{picture}(5,5)(0,0)
\put(0,0.5){\line(1,0){5}}
\put(0,1.5){\line(1,0){5}}
\put(0,2.5){\line(1,0){5}}
\put(0,3.5){\line(1,0){5}}
\put(0,4.5){\line(1,0){5}}
\put(0.5,0){\line(0,1){5}}
\put(1.5,0){\line(0,1){5}}
\put(2.5,0){\line(0,1){5}}
\put(3.5,0){\line(0,1){5}}
\put(4.5,0){\line(0,1){5}}
\put(2.5,2.5){\color{blue}\line(1,1){1}}
\put(2.5,2.5){\color{blue}\line(1,-1){1}}
\put(2.5,2.5){\color{blue}\line(-1,1){1}}
\put(2.5,2.5){\color{blue}\line(-1,-1){1}}
\put(2.5,2.5){\color{red}\line(1,2){1}}
\put(2.5,2.5){\color{red}\line(-1,2){1}}
\put(2.5,2.5){\color{red}\line(2,1){2}}
\put(2.5,2.5){\color{red}\line(2,-1){2}}
\put(2.5,2.5){\color{red}\line(1,-2){1}}
\put(2.5,2.5){\color{red}\line(-1,-2){1}}
\put(2.5,2.5){\color{red}\line(-2,1){2}}
\put(2.5,2.5){\color{red}\line(-2,-1){2}}
\put(0.5,0.5){\circle*{0.8}}
\put(2.5,0.5){\circle*{0.8}}
\put(4.5,0.5){\circle*{0.8}}
\put(0.5,4.5){\circle*{0.8}}
\put(2.5,4.5){\circle*{0.8}}
\put(4.5,4.5){\circle*{0.8}}
\put(0.5,2.5){\circle*{0.8}}
\put(2.5,2.5){\circle*{0.8}}
\put(4.5,2.5){\circle*{0.8}}
\end{picture}
\end{center}
\caption{One of the ground-state configurations for the interaction Hamiltonian $H_{\rm int}^{(\Lambda)}$ at the quarter filling. 
The sites of the solid circles are occupied by the fermions.}
\label{GSconfig}
\end{figure}

Let us consider the ground states of the interaction Hamiltonian $H_{\rm int}^{(\Lambda)}$ of (\ref{Hint}) 
with no hopping Hamiltonian. We choose $\Lambda=[1,4L]\times[1,4L]$ with a positive integer $L$ 
with the periodic boundary conditions, and consider the quarter filling. From the assumptions on the interactions, 
It is easily shown that the ground-state space is four-fold degenerate, and one of the ground-state configurations is depicted 
in Fig.~\ref{GSconfig}.

A reader might think that these ground states show a charge-density-wave long-range order \cite{LB} because 
the translational invariance is spontaneously broken under the unit lattice shift. 
However, in order to realize a fractional Chern insulator, we need a large structure of a unit cell 
with a larger period for the translation. As we will see below, this structure can be realized by 
a certain hopping Hamiltonian. 
The situation is very similar to those in the usual topological insulators. 
Therefore, the structure of the ground-state configuration should be interpreted as a microscopic structure 
in the unit cell. In fact, the above ground states do not break the translational invariance of the total 
Hamiltonian, which includes the hopping Hamiltonian.

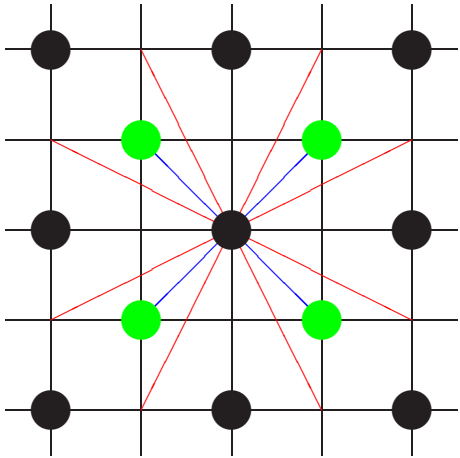
\begin{figure}[ht]
\setlength{\unitlength}{1.2cm}
\begin{center}
\begin{picture}(5,5)(0,0)
\put(0,0.5){\line(1,0){5}}
\put(0,1.5){\line(1,0){5}}
\put(0,2.5){\line(1,0){5}}
\put(0,3.5){\line(1,0){5}}
\put(0,4.5){\line(1,0){5}}
\put(0.5,0){\line(0,1){5}}
\put(1.5,0){\line(0,1){5}}
\put(2.5,0){\line(0,1){5}}
\put(3.5,0){\line(0,1){5}}
\put(4.5,0){\line(0,1){5}}
\put(2.5,2.5){\color{blue}\line(1,1){1}}
\put(2.5,2.5){\color{blue}\line(1,-1){1}}
\put(2.5,2.5){\color{blue}\line(-1,1){1}}
\put(2.5,2.5){\color{blue}\line(-1,-1){1}}
\put(2.5,2.5){\color{red}\line(1,2){1}}
\put(2.5,2.5){\color{red}\line(-1,2){1}}
\put(2.5,2.5){\color{red}\line(2,1){2}}
\put(2.5,2.5){\color{red}\line(2,-1){2}}
\put(2.5,2.5){\color{red}\line(1,-2){1}}
\put(2.5,2.5){\color{red}\line(-1,-2){1}}
\put(2.5,2.5){\color{red}\line(-2,1){2}}
\put(2.5,2.5){\color{red}\line(-2,-1){2}}
\put(0.5,0.5){\circle*{0.8}}
\put(1.5,1.5){\color{green}\circle*{0.8}}
\put(3.5,1.5){\color{green}\circle*{0.8}}
\put(1.5,3.5){\color{green}\circle*{0.8}}
\put(3.5,3.5){\color{green}\circle*{0.8}}
\put(2.5,0.5){\circle*{0.8}}
\put(4.5,0.5){\circle*{0.8}}
\put(0.5,4.5){\circle*{0.8}}
\put(2.5,4.5){\circle*{0.8}}
\put(4.5,4.5){\circle*{0.8}}
\put(0.5,2.5){\circle*{0.8}}
\put(2.5,2.5){\circle*{0.8}}
\put(4.5,2.5){\circle*{0.8}}
\end{picture}
\end{center}
\caption{Additional fermions must occupy the sites depicted by the green solid circles to lower the energy 
in the strong coupling limit $g_1\rightarrow\infty$.}
\label{GSconfig2}
\end{figure}

Further, by adding fermions, we want to realize the filling $1/4+1/8=3/8$. 
For this purpose, we assume $g_1\gg g_2>0$ for the two coupling constants of the interactions. 
{From} these assumptions, the additional fermions must occupy the sites depicted by 
the green solid circles in Fig.~\ref{GSconfig2} to lower the energy.

\Section{Hopping Hamiltonians}
\label{Sec:hopHam}

Now, let us take into account the hopping of the fermions. We consider the strong coupling limit $g_1\rightarrow\infty$ 
for the Hamiltonian $H_{{\rm int},1}^{(\Lambda)}$ of (\ref{Hint1}). In this limit, it is enough to 
consider the hopping between the sites depicted by the green solid circles in Fig.~\ref{GSconfig2}. 
For example, the strong nearest-neighbor Coulomb repulsion indeed forbids the nearest-neighbor hopping.  
Clearly, these green sites form the square lattice for each classical ground state which is depicted by 
the black solid circles. 

For this square lattice, we consider the hopping model which was introduced by 
Hatsugai and Kohmoto \cite{HK}. See Appendix~\ref{HKmodel} for the details of the model. 
In the case of the half-filled band, the model shows the Chern number $\pm 1$, depending on the model parameters. 
Therefore, for each of the four ground states of the interaction Hamiltonian 
$H_{{\rm int},1}^{(\Lambda)}$ of (\ref{Hint1}), there are two choices of the Chern numbers. 
We denote a set of the four Chern numbers by $(\sigma_1,\sigma_2,\sigma_3,\sigma_4)\in\{+1,-1\}^4$. 
 
Consider first the case of $(+1,+1,+1,-1)$ for the Chern numbers, i.e., the three are equal to $+1$, 
and the rest is equal to $-1$. In this case, the Chern number of the total system is given by 
\begin{equation}
\mbox{Chern number}=\frac{1}{4}(1+1+1-1)=\frac{1}{2}
\end{equation}
in the sense of the expectation value about the four-degenerate ground state \cite{NTW}. 
Thus, the Hall conductance is quantized to the fractional value $1/2$. For this choice of the set of the four Chern numbers, 
the corresponding total hopping Hamiltonian is not invariant under the unit lattice translation
in both of the two spatial directions. Actually, the model parameters for one of the four hopping Hamiltonians of Hatsugai 
and Kohmoto must be chosen to be different from those for the remaining three hopping Hamiltonians 
so that they realize $(+1,+1,+1,-1)$ for 
the set of the four Chern numbers.   
Therefore, all the four ground states of the interaction Hamiltonian 
$H_{{\rm int},1}^{(\Lambda)}$ of (\ref{Hint1}) do not break the translational symmetry of the total Hamiltonian 
which consist of the interaction and the hopping Hamiltonians. 
In this sense, one cannot interpret any of these four ground states as a charge-density-wave(CDW) state. 
However, the four ground states of the interaction Hamiltonian $H_{{\rm int},1}^{(\Lambda)}$ of (\ref{Hint1}) 
look like to exhibit a long-range order of charge density wave. 
Since the strong coupling limit $g_1\rightarrow\infty$ is very special,  
it may change due to the effect of the asymmetry of the four two-site hopping Hamiltonians 
for a sufficiently large but finite coupling constant $g_1$ and sufficiently small but non-zero hopping amplitudes  
of a nearest-neighbor hopping Hamiltonian.

\Section{CDW versus FCI}
\label{Sec:CDWFCI}

We can also choose the four Chern numbers to be all the same, e.g., $(+1,+1,+1,+1)$,  
so that the system is invariant under the unit lattice translation in one of the spatial directions. 
In this case, the four ground states break the translational invariance of the total Hamiltonian. 
Therefore, the long-range order of the charge density wave emerges.  
On the other hand, the Chern number is given by the integer $(1+1+1+1)/4=1$. 
Clearly, this phase can be distinguished from that in the preceding section by using the Chern numbers. 
Thus, this phase can be characterized by the long-range order of the charge-density wave with the Chern number $+1$.

\appendix 

\Section{Hatsugai-Kohmoto hopping model}
\label{HKmodel}

In this appendix, we present a short review about the tight-binding model by Hatsugai and Kohmoto \cite{HK} on the square lattice. 

\begin{figure}[ht]
\setlength{\unitlength}{1.5cm}
\begin{center}
\begin{picture}(5,3)(0,0)
\put(0,0.5){\line(1,0){5}}
\put(0,2.5){\line(1,0){5}}
\put(0.5,0){\line(0,1){3}}
\put(2.5,0){\line(0,1){3}}
\put(4.5,0){\line(0,1){3}}
\put(0,0){\line(1,1){3}}
\put(2,0){\line(1,1){3}}
\put(3,0){\line(-1,1){3}}
\put(5,0){\line(-1,1){3}}
\put(1.5,0.2){$t_1$}
\put(0.8,1.5){$it_{\rm d}$}
\put(1.3,1.5){\vector(-1,1){0.4}}
\put(1.18,0.75){$it_{\rm d}$}
\put(0.95,0.75){\vector(1,1){0.4}}
\put(3.2,0.75){$it_{\rm d}$}
\put(3.35,1.14){\vector(-1,-1){0.4}}
\put(2.8,1.5){$it_{\rm d}$}
\put(2.9,1.9){\vector(1,-1){0.4}}
\put(3.5,0.2){$t_1$}
\put(1.5,2.2){$t_1$}
\put(3.5,2.2){$t_1$}
\put(0.2,1.5){$t_2$}
\put(1.95,1.5){$-t_2$}
\put(4.2,1.5){$t_2$}
\end{picture}
\end{center}
\caption{The hopping amplitudes of the tight-binding model on the square lattice \cite{HK}. 
The diagonal hopping amplitudes are pure imaginary, and hence they depend on the direction of the hopping.}
\label{HKmodelFig}
\end{figure}
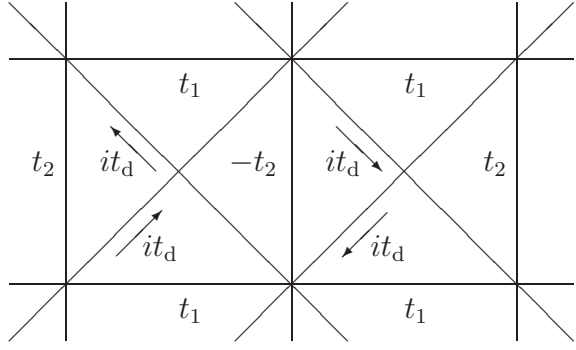

Let us consider the $2L_1\times L_2$ square lattice, where $L_1$ and $L_2$ are a positive integer. 
We write $\varphi(\ell,n)$ for the single-particle wavefunction at the site $(\ell,n)$, 
$\ell=1,2,\ldots,2L_1$ and $n=1,2,\ldots,L_2$, with the periodic boundary conditions. 
The hopping amplitudes of Hatsugai-Kohmoto model are depicted in Fig.~\ref{HKmodelFig}.  

The Schr\"odinger equation for the single-particle wavefunction is given by 
\begin{eqnarray}
& &t_1[\varphi(2m-2,n)+\varphi(2m,n)]+t_2[\varphi(2m-1,n-1)+\varphi(2m-1,n+1)]\nonumber\\
&+&it_{\rm d}[-\varphi(2m-2,n-1)-\varphi(2m,n+1)+\varphi(2m,n-1)+\varphi(2m-2,n+1)]\nonumber\\
&=&E\varphi(2m-1,n)
\end{eqnarray}
and 
\begin{eqnarray}
& &t_1[\varphi(2m-1,n)+\varphi(2m+1,n)]-t_2[\varphi(2m,n+1)+\varphi(2m,n-1)]\nonumber\\
&+&it_{\rm d}[\varphi(2m-1,n-1)+\varphi(2m+1,n+1)-\varphi(2m+1,n-1)-\varphi(2m-1,n+1)]\nonumber\\
&=&E\varphi(2m,n)
\end{eqnarray}
for $m=1,2,\ldots,L_1$, and $n=1,2,\ldots,L_2$. Here, $E$ is the energy eigenvalue, $t_1,t_2\in\re$, and we have chosen 
the same value $t_{\rm d}\in\re$ for the two diagonal hopping amplitudes \cite{HK} for simplicity. 
The model is invariant under the translation by two lattice units in the first direction and by one lattice unit 
in the second direction. 

Because of the translational invariance, we can take the wavefunction in the plane-wave form, 
\begin{equation}
\varphi(2m-1,n)=\alpha_k \exp[ik^{(1)}m+ik^{(2)}n] \quad \mbox{and}\quad 
\varphi(2m,n)=\beta_k \exp[ik^{(1)}m+ik^{(2)}n],
\end{equation}
with coefficients, $\alpha_k$ and $\beta_k$, and the wave vector $k=(k^{(1)},k^{(2)})$. 
Substituting these forms of the wavefunction into the Schr\"odinger equation, one has 
\begin{equation}
\begin{pmatrix}
A(k) & B(k) \\
B^\ast(k) & -A(k) 
\end{pmatrix}
\begin{pmatrix}
\alpha_k \\ \beta_k
\end{pmatrix}
=E
\begin{pmatrix}
\alpha_k \\ \beta_k
\end{pmatrix},
\end{equation}
where 
\begin{equation}
\label{A(k)}
A(k):=2t_2\cos k^{(2)}
\end{equation}
and 
\begin{equation}
\label{B(k)}
B(k):=e^{-ik^{(1)}/2}\left[2t_1\cos \frac{k^{(1)}}{2}+4i t_{\rm d}\sin \frac{k^{(1)}}{2}\sin k^{(2)}\right].
\end{equation}
The two energy eigenvalues are given by 
\begin{equation}
E_k:=\pm \sqrt{[A(k)]^2+|B(k)|^2}.
\end{equation}
The eigenvector with the low energy eigenvalue is 
\begin{equation}
\begin{pmatrix}
\alpha_k \\ \beta_k
\end{pmatrix}
=
\begin{pmatrix}
B(k) \\ -A(k)-\sqrt{[A(k)]^2+|B(k)|^2}
\end{pmatrix}.
\end{equation}

In order to calculate the Chern number for the low band, we first examine a zero point 
of the norm $N_k:=\sqrt{|\alpha_k|^2+|\beta_k|^2}$ of the eigenvector. Clearly, one has 
\begin{equation}
N_k^2=|B(k)|^2+\left\{A(k)+\sqrt{[A(k)]^2+|B(k)|^2} \right\}^2.
\end{equation}
As is well known, a wavenumber $k$ which yields the vanishing of the norm $N_k$ leads to a non-trivial contribution 
for the Chern number. The point is determined by 
\begin{equation}
|B(k)|=0 \quad \mbox{and}\quad A(k)+|A(k)|=0.
\end{equation}
{From} this first equation and the expression (\ref{B(k)}) of $B(k)$, the point must satisfy 
\begin{equation}
k^{(1)}=\pi \quad \mbox{and} \quad k^{(2)}= 0 \ \mbox{or} \ \pi .
\end{equation}
Further, from the expression (\ref{A(k)}) of $A(k)$ and these observations, the second component $k^{(2)}$ of the point 
must satisfy  
\begin{equation} 
k^{(2)}=
\begin{cases}
\pi & \mbox{for \ } t_2>0; \\
0  & \mbox{for \ } t_2<0.
\end{cases}
\end{equation}
In the following, we assume $t_2>0$. Therefore, if $N_k=0$, then $k=(\pi,\pi)$. 

Let us compute the Chern number which is given by 
\begin{equation}
{\rm Chern \ number}:=\frac{1}{2\pi i}\int_{-\pi}^\pi dk^{(1)}\int_{-\pi}^\pi dk^{(2)}\; {\rm rot}\;\mathcal{A}(k),
\end{equation}
where the two-component vector $\mathcal{A}(k)$ is defined by 
\begin{equation}
\mathcal{A}(k):=\frac{1}{N_k}(\alpha_k^\ast,\beta_k^\ast)\nabla_k
\frac{1}{N_k}\begin{pmatrix}
\alpha_k \\ \beta_k 
\end{pmatrix}.
\end{equation}
As is well known, the Chern number can be expressed as 
\begin{equation}
{\rm Chern \ number }=\lim_{\varepsilon \searrow 0}\frac{1}{2\pi i}\oint_{\gamma_\varepsilon} dk \cdot \mathcal{A}(k), 
\end{equation}
where $\gamma_\varepsilon$ is a small closed path which encircles the above point $k=(\pi,\pi)$. 

In order to calculate the above line integral, we set 
\begin{equation}
(k^{(1)},k^{(2)})=(\pi+p^{(1)},\pi+p^{(2)})
\end{equation}
with small $p^{(1)}, p^{(2)}$. Substituting this into the expression of $\mathcal{A}(k)$, one has 
\begin{equation}
\mathcal{A}(k)\approx 
i\begin{pmatrix}
-{1}/{2} \\ 0 
\end{pmatrix}
+\frac{4it_{\rm d}}{t_1}\frac{1}{(p^{(1)})^2 +\left({4t_{\rm d}}/{t_1}\right)^2(p^{(2)})^2}
\begin{pmatrix}
-p^{(2)} \\ p^{(1)}
\end{pmatrix},
\end{equation}
where we have assumed $t_1\ne 0$. The first term in the right-hand side does not contribute to the line integral. 
In order to compute the contribution by the second term in the line integral, we choose 
\begin{equation}
(p^{(1)},p^{(2)})=\left(\varepsilon \cos\theta, \varepsilon\left|\frac{t_1}{4t_d}\right|\sin\theta\right), \ \theta\in[0,2\pi)
\end{equation}
as the closed path $\gamma_\varepsilon$ with the small $\varepsilon>0$. Here, we have assumed $t_{\rm d}\ne 0$. 
In consequence, one obtains 
\begin{equation}
\mbox{Chern number}={\rm sgn}(t_{\rm d}/t_1),  
\end{equation}
where ${\rm sgn}$ stands for the signature.

\bigskip\bigskip


\end{document}